\documentclass[usenatbib,useAMS]{mn2e}
\usepackage{epsfig}
\usepackage{times}
\usepackage{amsmath}

\usepackage{verbatim}

\newcommand{\simgt}%
        {\,\hbox{\lower0.6ex\hbox{$\sim$}\llap{\raise0.6ex\hbox{$>$}}}\,}
\newcommand{\simlt}%
        {\,\hbox{\lower0.6ex\hbox{$\sim$}\llap{\raise0.6ex\hbox{$<$}}}\,}
\newcommand{\mytilde}{\raise.17ex\hbox{$\scriptstyle\mathtt{\sim}$}}
\usepackage{color}
\title[]{Synthetic X-ray spectra for simulations of the dynamics of an accretion flow irradiated by a quasar}
\author[]{S. A. Sim$^1$, D. Proga$^{2,3}$, R. Kurosawa$^4$, K. S. Long$^5$, L. Miller$^6$, T. J. Turner$^7$\\
$^{1}$Research School of Astronomy and Astrophysics, Mount Stromlo Observatory,
Cotter Road, Weston Creek, ACT 2611, Australia\\ 
$^{2}$Department of Physics, University of Nevada, Las Vegas, NV 89154, USA\\
$^{3}$Princeton University Observatory, Peyton Hall, Princeton, NJ 08540, USA\\
$^{4}$Department of Astronomy, Cornell University, Space Sciences Building, Ithaca, NY 14853-6801, USA\\
$^{5}$Space Telescope Science Institute, 3700 San Martin Drive, Baltimore, MD 21218, USA\\
$^{6}$Department of Physics, University of Oxford, Denys Wilkinson Buiding, Keble Road, Oxford OX1 3RH\\
$^{7}$Department of Physics, University of Maryland Baltimore County, 1000 Hilltop Circle, Baltimore, MD 21250, USA
}

\date{\today}

\pagerange{\pageref{firstpage}--\pageref{lastpage}}
\pubyear{}
\begin{document}
\maketitle
\label{firstpage}

\begin{abstract}
Ultraviolet and X-ray observations show evidence of outflowing gas around many active galactic nuclei. It has been proposed that some of these outflows are driven off gas infalling towards the central supermassive black hole.
We perform radiative transfer calculations to compute the gas ionization state 
and the emergent X-ray spectra for both two- and three-dimensional (3D) 
hydrodynamical simulations of this outflow-from-inflow scenario. By comparison 
with observations, our results can be used to test the theoretical models and
guide future numerical simulations.
We predict 
both absorption and emission features, most of which are
formed in a polar funnel of relatively dense ($10^{-20}$ -- $10^{-18}$~g~cm$^{-3}$) outflowing gas.
This outflow causes strong absorption for observer orientation angles of $\simlt 35$~degrees. Particularly in 3D, the strength
of this absorption varies significantly for different lines-of-sight
owing to the fragmentary structure of the gas flow. 
Although infalling material occupies a large fraction of the simulation volume, we do not
find that it imprints strong absorption features in the X-ray spectra since the ionization state is predicted to be very high. Thus, an 
absence of observed inflow absorption features does not exclude the models. The main spectroscopic consequence of the infalling gas is
a Compton-scattered continuum 
component that partially re-fills the absorption features caused by the outflowing 
polar funnel.
Fluorescence and scattering in the outflow is predicted to give rise to 
several emission features including a multi-component 
Fe~K$\alpha$ emission complex
for all observer orientations. 
For the hydrodynamical simulations considered, 
we predict both ionization states and column densities 
for the outflowing gas that are too high to be quantitatively consistent with well-observed X-ray absorption systems.
Nevertheless, our results are qualitatively encouraging and further exploration of the model 
parameter space is warranted. 
Higher resolution hydrodynamic simulations are needed to determine
whether the outflows fragment on scales unresolved in our current study, which
may yield the denser lower ionization material that could reconcile the
models and the observations.
\end{abstract}


\section{Introduction}
\label{sect_intro}

The energy liberated during accretion by supermassive black holes can be sufficient that active galactic nuclei (AGN) could significantly influence
their environments. Indeed, it has been argued that feedback from AGN may play an 
important role in the evolution of galaxies \citep[e.g.][]{croton06}. 
One potential mechanism by which AGN may affect their host galaxy is via energy and momentum carried by winds or outflows \citep{king03b,king05,ciotti07,ciotti09,ciotti10,ostriker10}.

There is copious observational evidence suggesting that many AGN have associated outflows. In particular, blue-shifted 
absorption lines are commonly observed
in both ultraviolet (uv) and X-ray spectra \citep[see reviews by][]{crenshaw03,turner09}. These absorption lines suggest 
outflows exist with a wide variety of 
properties: although the shifts of many uv and soft X-ray lines typically suggest outflow speeds of only a few hundreds to thousands of 
km~s$^{-1}$, uv spectra of broad absorption line quasars show line shifts up to $\sim 0.1$c \citep{weymann81} and evidence of 
even higher velocity flows has been found in X-ray observations \citep[e.g.][]{reeves03,chartas02,tombesi10}. 
In addition, detection of rapid ($\sim$ ks) 
variability in some highly ionized absorption components implies that that they are located relatively close ($\sim$ hundreds of gravitational radii) 
to the AGN \citep{reeves04, miller07,turner07,turner11}. In other cases, however, lack of variability has been used to argue that components of the X-ray absorbing gas are distant from the AGN, likely located on parsec scales from the central black hole \citep{netzer03,behar03,krongold10,kaastra12}. This wide diversity of absorber properties suggests that there may be multiple means by which AGN drive outflows and so theoretical modelling for a variety of outflow mechanisms is required to build a complete picture.

Since AGN are extremely luminous, it is natural to investigate the role of radiation in ejecting material from their environment.
In a previous study \citep{sim10c}, we computed synthetic X-ray spectra for simulations of a line-driven
AGN accretion disk wind \citep{proga04} launched from within a few hundred gravitational radii of a supermassive black hole. 
We found that such an outflow might account for a variety of observable spectral features associated with 
highly ionized, high-velocity gas. However, many observed spectral features (e.g. absorption lines with relatively small blueshifts or narrow 
components of emission lines) are unlikely to be formed in outflows very close to the central black hole. These are more probably associated
with distant material where characteristic flow velocities are expected to be smaller. 
One promising origin for these distant outflows is a wind blown from the inner regions of a torus of dusty material surrounding the AGN \citep{balsara93,dorodnitsyn08a,dorodnitsyn08b,dorodnitsyn09}. In that model, the mass reservoir for the outflow is rotationally supported cold material and the resulting outflow has velocities of $\sim 1000$~km~s$^{-1}$ close to the rotation axis, dropping to a few hundred km~s$^{-1}$ at higher inclinations \citep{dorodnitsyn08b}. Such a model predicts observable outflow features which are broadly compatible with the properties of many observed X-ray absorbers. 

An alternative model is that in which an outflow is driven not from a rotationally supported torus but directly from gas infalling towards
a supermassive black hole. This scenario has been modelled on parsec scales by 
\cite{proga07}, \cite{proga08}, Kurosawa \& Proga (2009b, hereafter KP09)
\nocite{kurosawa09}
and \cite{kurosawa09b}.
It was found that momentum and energy supplied by radiation from the AGN could strongly affect the dynamics of infalling gas, giving rise to a characteristic flow pattern which includes a polar outflow driven off the infalling gas. \cite{proga07} and \cite{proga08} studied this process using two-dimensional (2D) simulations both with and without gas rotation and stressed that the formation of such a flow pattern was probable if the AGN luminosity is larger than around one per cent of the Eddington luminosity. KP09 extended the studies to included fully three-dimensional (3D) simulations and confirmed that the characteristic flow pattern of the earlier 2D calculations was also robust in 3D. \cite{kurosawa09b} examined the
consequences of coupling the black hole accretion luminosity to the mass inflow rate in the simulations and still found the same flow pattern for a outflow radiatively driven off hot infalling gas in many cases.

The goal of this paper is to compute synthetic X-ray spectra by carrying out radiative transfer simulations based on the KP09 hydrodynamical calculations. These synthetic spectra will allow us to quantify the observable consequences of the outflow/inflow pattern and asses how well they might account for observed spectroscopic features. 
One key motivation of this work is the question of whether the {\it infalling} gas in the simulation strongly affects the X-ray spectra. Signatures of infall are not commonly observed and so it
 is important
to consider whether this is compatible with models for outflows that are driven off a mass reservoir of infalling gas. We also wish to compare the observational features predicted by
3D outflow simulations with those of axi-symmetric 2D calculations. KP09 showed that the overall flow pattern in 3D was similar to that in 2D but with significantly more structure in the
polar outflow. Our radiative transfer simulations allow us to quantity the observable
consequences of this extra complexity.

We begin by summarising the key properties of the KP09 simulations and our radiative transfer calculations in Section~\ref{sect:sims}. We then present our synthetic spectra in Section~\ref{sect:results} and discuss their interpretation in Section~\ref{sect:discuss}. Our conclusions are summarised in Section~\ref{sect:summary}.

\begin{figure*}
\epsfig{file=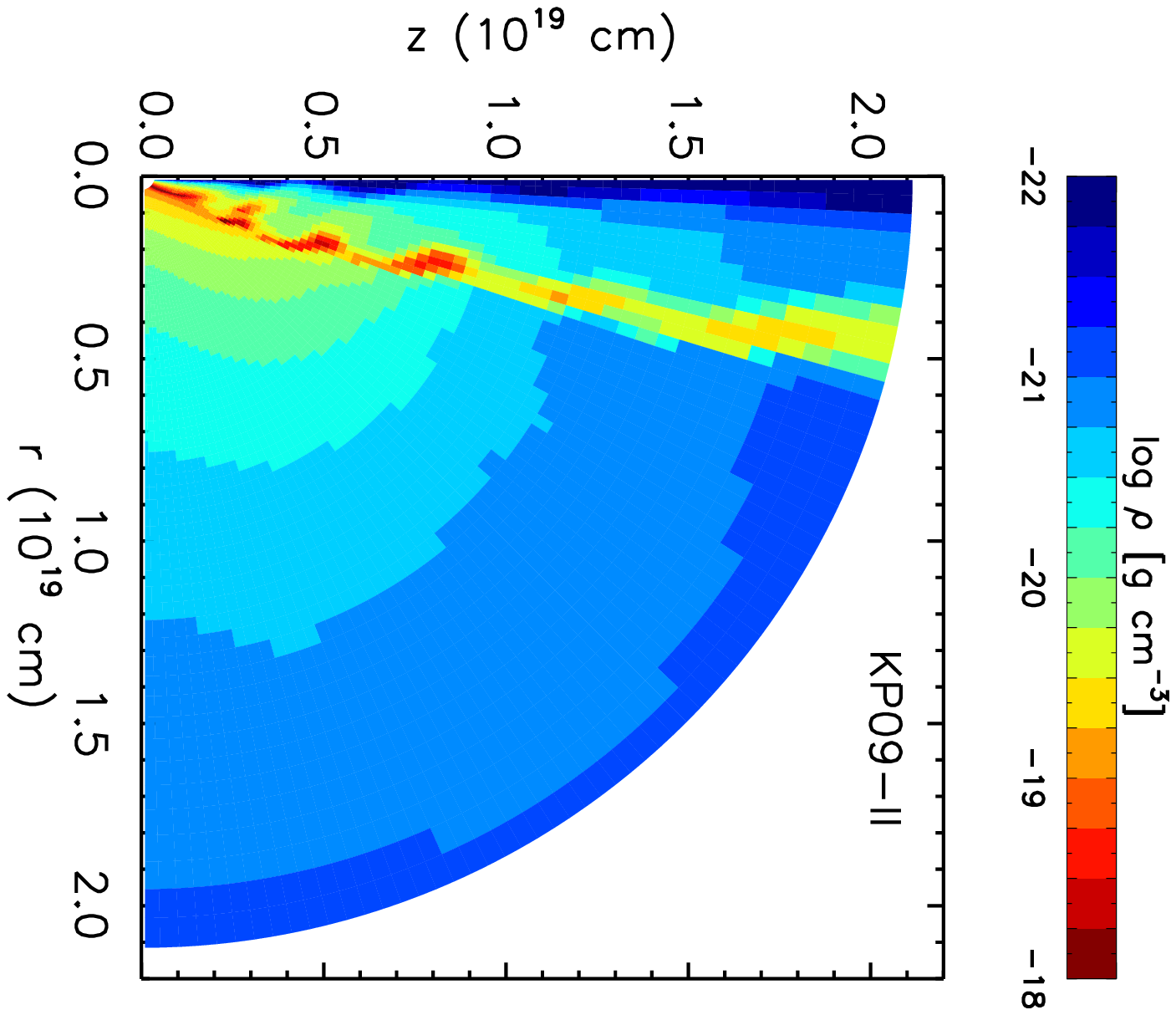,width=8cm,angle=90}
\epsfig{file=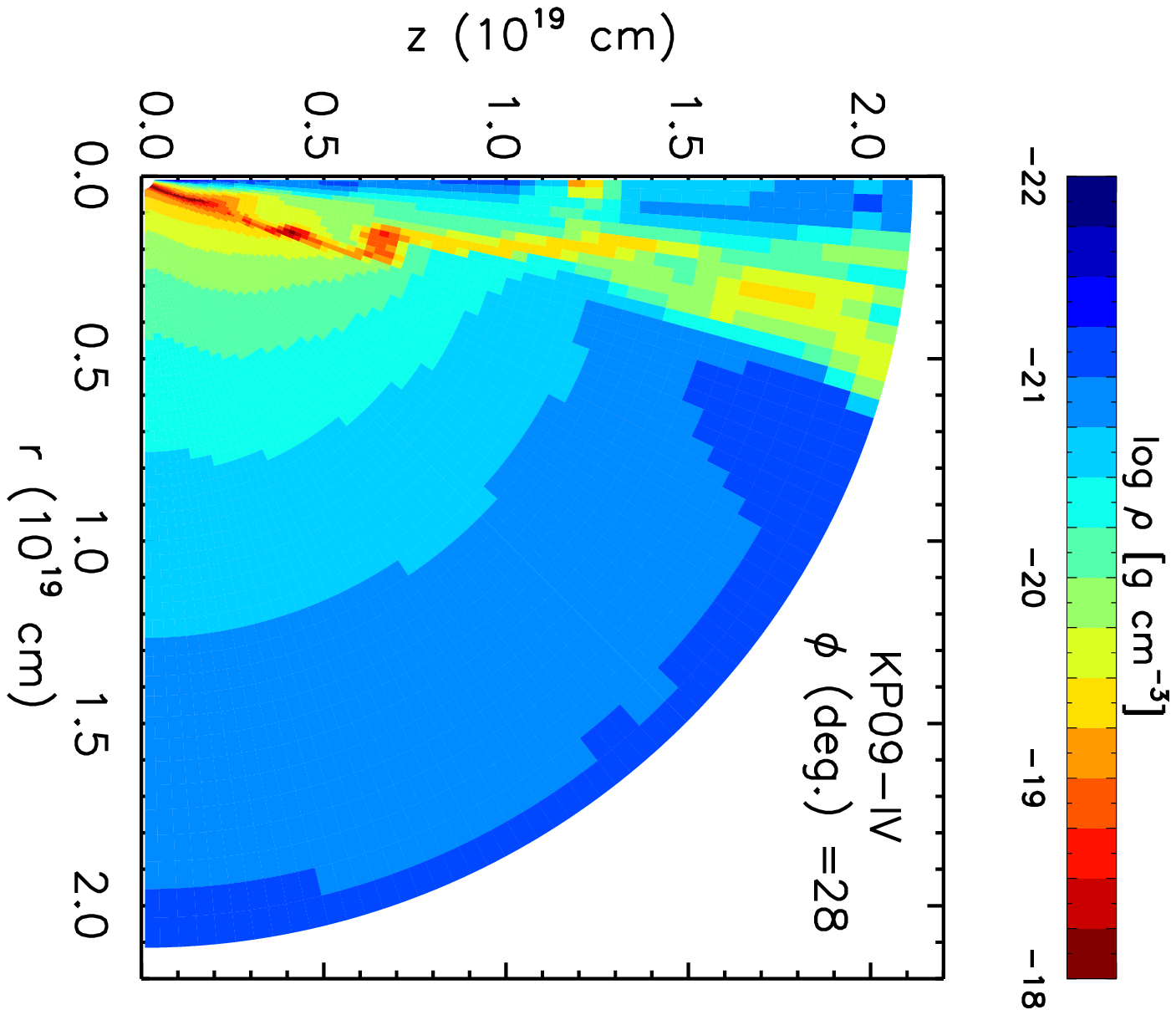,width=8cm,angle=90}\\
\epsfig{file=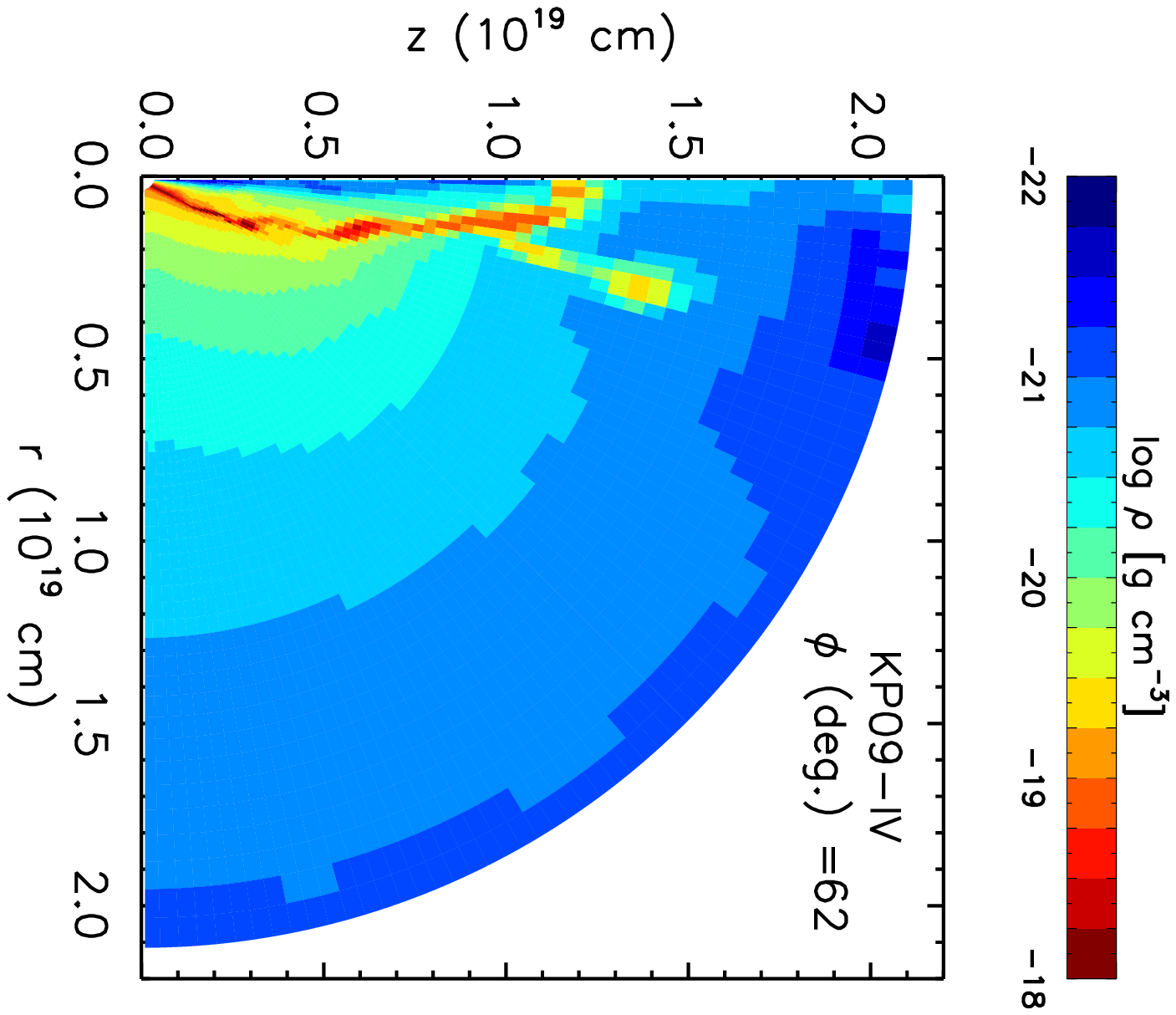,width=8cm,angle=90}
\epsfig{file=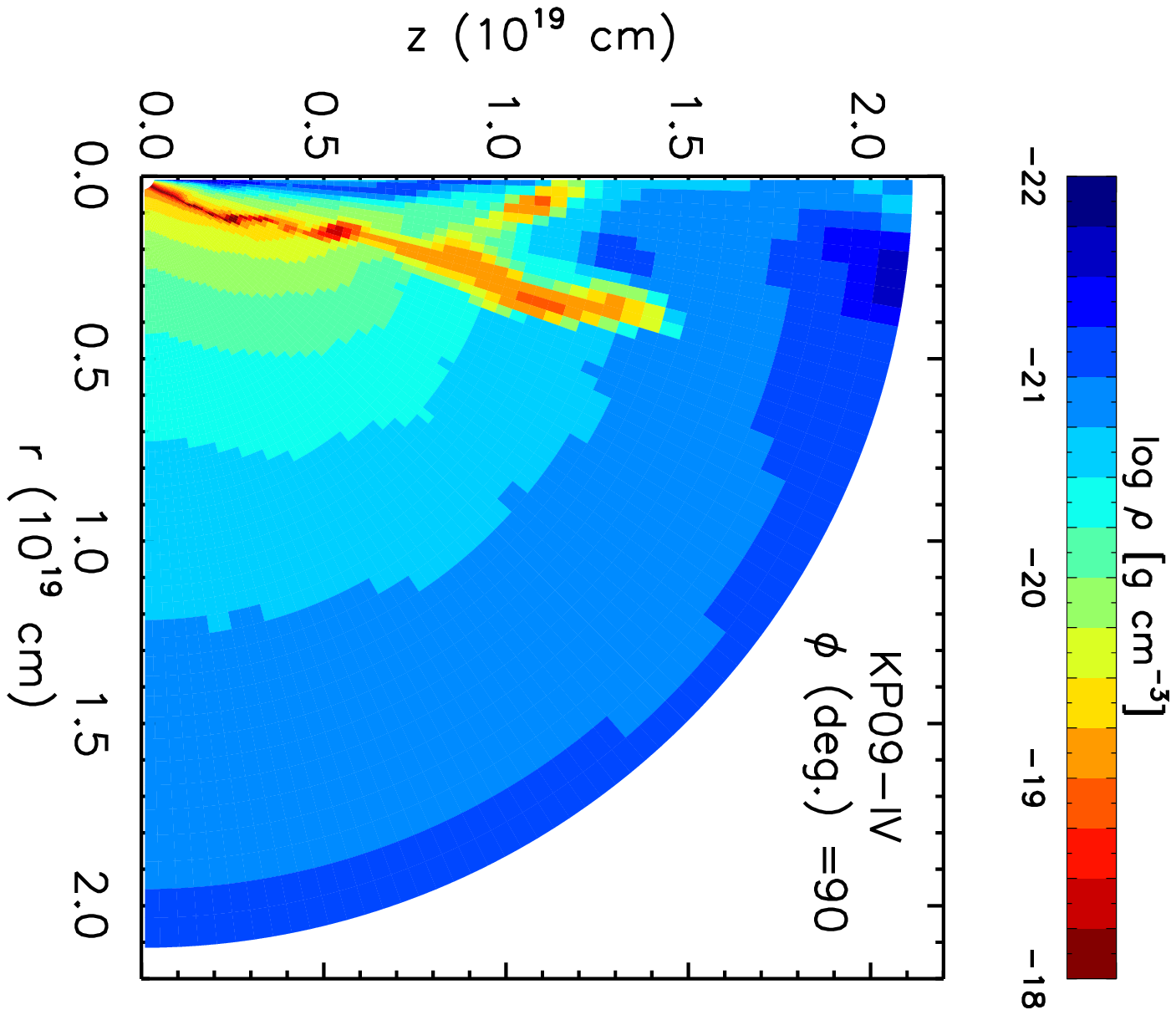,width=8cm,angle=90}\\
\caption{Density distributions in the $+z$ region for the models we consider. The top left panel show a slice in the $x=0$-plane for our two-dimensional model (KP09-II), which is symmetric under rotation about the $z$-axis. The remaining three panels show slices through the density distribution of our three-dimensional model (KP09-IV) in the planes defined by the spherical polar coordinate $\phi = 28$, 62 and 90~degrees.}
\label{fig:rho_plots}
\end{figure*}

\section{Numerical simulations}
\label{sect:sims}

\subsection{Input from radiation hydrodynamical simulations}

As input for our radiative transfer calculations, we have taken 
snapshots of density and velocity from two of the
radiation hydrodynamical simulations presented by KP09.
Their simulations are for the region $3000 < r/r_{\rm g} < 1.5 \times 10^6$ (where $r_{\rm g} = G M_{\rm BH} / c^2 = 1.5 \times 10^{13}$~cm) around a central black hole of mass $M_{\rm BH} = 10^{8}$~$M_{\odot}$. An accretion disk surrounding the black hole is assumed to radiate at a bolometric luminosity of $L_{\rm bol} = 7.6 \times 10^{45}$~ergs~s$^{-1}$.
The simulations were initialized with a uniform mass density ($\rho_0 = 10^{-21}$~g~cm$^{-3}$) and gas temperature ($T_o = 2 \times 10^7$~K). Radial velocities were initialized to zero but rotational motions were included in the setup as described in section 2.3 of KP09. The polar axis is treated with an axis-of-symmetry boundary condition while outflow boundary conditions are applied at both
the inner and outer radial boundaries. The density and temperature of material entering through the outer boundary of the simulation (at $r = 1.5 \times 10^6 r_{\rm g} \approx 7$~pc) were fixed to $\rho_0$ and $T_{0}$ while the radial velocity at the 
boundary was allowed to float
(see KP09 for full details of the simulation setup\footnote{For simulation movies, see\\ http://www.astro.cornell.edu/\mytilde kurosawa/research/agn.html.}).

We have studied both the KP09 simulations that included gas rotation: their axi-symmetric simulations (model II) and full three-dimensional simulation (model IV). The initial conditions for these two hydrodynamical simulations were identical -- the only difference was the assumed axi-symmetry in model II. Thus, comparison of these simulations will allow us to quantify how the imposed axi-symmetry of two-dimensional models influences our conclusions. We will refer to these models as KP09-II and KP09-IV, respectively. 
In physical units, the time between the snapshots recorded during the KP09 hydrodynamical simulations was $\sim 300$~yr. This is too long for it to be useful to compare multiple snapshots to estimate how the spectral features may vary over observable timescales. Consequently we consider only one representative snapshot for each simulation.
The particular snapshots we adopt are taken from the phase where the hydrodynamical simulation has evolved into a time-averaged steady state and are 
the same as those shown in figure 2 of KP09. 
Neither the hydrodynamical simulations nor our radiative transfer calculations assume 
symmetry about the $xy$-plane (i.e. the full interval for the spherical 
polar angle is included, $0 < \theta < 180$~deg.). However, the flow structure
is very similar in the $+z$ and $-z$ hemispheres and so we will focus our 
discussion on only the $+z$ hemisphere ($0 < \theta < 90$~deg); in all 
important respects, the $-z$ region mirrors the same behaviour.

Figure~\ref{fig:rho_plots} shows two-dimensional slices of the density in the model snapshots we have studied 
(for KP09-IV we include three representative slices). 
In general, these all show similar features, as described by KP09. The flow structure can roughly be divided into three regimes. Very close to the rotation axis ($\theta \simlt 15$~deg), there is a region of very low density gas ($\rho < 10^{-21}$~g~cm$^{-3}$), which is flowing away from the central source. Surrounding this is an outflowing funnel of relatively dense gas (typically $\rho > 2 \times 10^{-20}$~g~cm$^{-3}$) which is being accelerated out of the system (the total mass outflow rate through the outer boundary of the simulation is $\sim 0.8$~M$_{\odot}$~yr$^{-1}$). 
The gas in the polar funnel is thermally unstable. The hot thermally stable infalling gas in the funnel is brought to the regime of thermal instability by the action of the centrifugal force and radiation pressure on electrons -- both forces slow down the gas and lead to a density increase and fast cooling. This in turn enhances the outward line force that is sensitive to the density, photoionization parameter and optical depth. Consequently, a fragmented outflow develops with
dense substructures (exceeding $\rho = 10^{-18}$~g~cm$^{-3}$ in places). 
Outside this outflowing stream ($\theta \simgt 35$~deg), the gas density is significantly lower (in most of this region, $\rho < 10^{20}$~g~cm$^{-3}$) and the predominant motion is infall towards the central source
(the net inflow rate through the inner boundary is $\sim 0.8$~M$_{\odot}$~yr$^{-1}$).
The infalling gas outside the funnel is supersonic, heated mostly by adiabatic compression and its temperature is high so that this gas is always thermally stable.

\begin{figure}
\epsfig{file=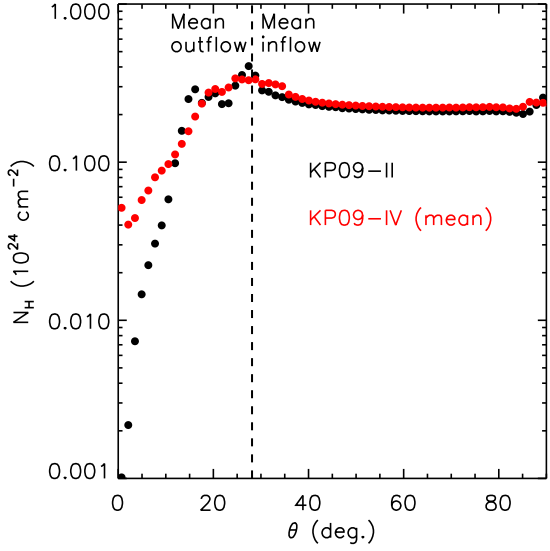,width=8cm,angle=90}
\caption{Hydrogen column density measured through the models from the origin to an observer at inclination $\theta$ to the $z$-axis for our models. For the three-dimensional KP09-IV model, we plot the column-density averaged over all azimuthal angles. 
The dashed vertical line marks the divide between orientations for which the KP09-II column-density weighted mean radial velocity is positive (i.e. the mean motion is outflow) and negative (i.e. mean inflow).
We show only the $+z$ hemisphere -- the same behaviour is mirrored in the $-z$ hemisphere.}
\label{fig:columns}
\end{figure}

The outflowing material is sufficiently dense that it can be expected to imprint strong signatures on the spectrum of the source. Figure~\ref{fig:columns} shows the hydrogen column density from the origin through our models as a function of inclination-angle ($\theta$, measured from the $z$-axis). For the KP09-IV model we show the column density averaged over all azimuthal angles. In model KP09-II, the column density is relatively modest ($N_{\rm H} < 10^{22}$~cm$^{-2}$) very close to the polar axis but is significant ($\sim 10^{23}$~cm$^{-2}$) for most lines-of-sight through the flow. The highest column densities ($\simgt 3 \times 10^{23}$~cm$^{-2}$) are for lines-of-sight that pass through the outflowing polar funnel ($15 < \theta < 35$~deg). We also note that the variation of the column-density with orientation is most complex in this region. This is caused by the fragmentation of the outflowing gas -- the very highest column densities are for lines-of-sight that pass through dense clouds. It is therefore to be expected that the most complex spectral features will be imprinted when the central X-ray source is viewed through this gas.

\begin{figure}
\epsfig{file=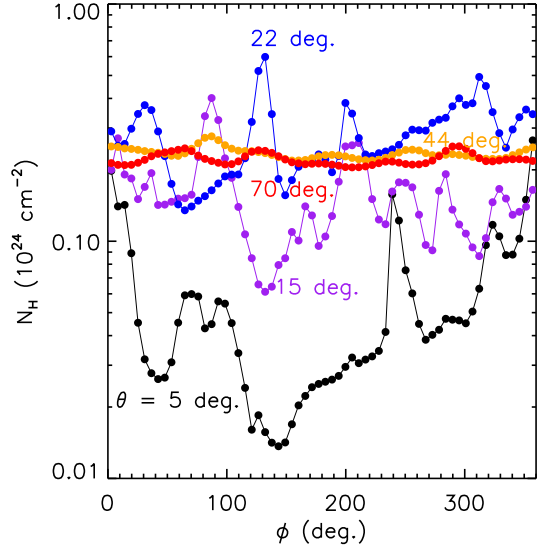,width=8cm,angle=90}
\caption{Hydrogen column density through model KP09-IV as a function of azimuthal angle ($\phi$) for fixed polar angle ($\theta$). Values are shown for five $\theta$-values: 5, 15, 22, 44 and 70~degrees (see labels in the Figure).}
\label{fig:columns_phi}
\end{figure}

In model KP09-IV, the density distribution is similar to that in KP09-II except that the outflowing funnel is now able to fragment into three-dimensional structures (i.e. there is no imposed axi-symmetry). Consequently, the density becomes a strong function of the azimuthal angle in the vicinity of the polar outflow. This is illustrated in Figure~\ref{fig:columns_phi}, which shows the variation of the column-density through our three-dimensional model (KP09-IV) as a function of azimuthal angle ($\phi$) for several fixed values of the polar angle ($\theta$). For moderate and high inclinations ($\theta > 40$~deg), the azimuthal variation of the column density is small. At these inclinations, the line-of-sight is mostly passing through the large volume of low-density, infalling material, which is not strongly fragmented. However, for smaller values of $\theta$, the variation of $N_{\rm H}$ with $\phi$ becomes increasingly significant -- for $\theta \sim 5$~deg., there is almost an order of magnitude of variation. 

\subsection{Radiative transfer simulations}

We have carried our radiative transfer calculations
for the snapshots from the KP09 simulations described above using our multi-dimensional Monte Carlo radiative transfer code \citep{sim05b,sim08,sim10b}.
Except where noted below, 
these simulations were performed in an identical manner to that described in \cite{sim10c}: we first carry out an iterative sequence of calculations to compute the ionization state of the gas assuming a centrally concentrated primary X-ray source. We then use our Monte Carlo quanta to construct detailed synthetic X-ray spectra for a range of observer orientations.

\begin{table*}
\caption{Parameters for the simulations.}
\label{tab:parameters}
\begin{tabular}{lc}
  \hline
  Parameter & Adopted value \\ \hline
Mass of central black hole, $M_{\rm bh}$ & $10^8$~M$_{\odot}$\\
  Bolometric luminosity, $L_{\rm bol}$ & $7.6 \times 10^{45}$~ergs~s$^{-1}$ 
($\sim 0.6 L_{\rm Edd}$) 
\\
  X-ray source luminosity (2 -- 10 keV), $L_{\rm X}$ & $1.5 \times 10^{44}$~ergs~s$^{-1}$ 
\\
  X-ray source power-law photon index, $\Gamma$ & 2.1\\
Range of source photons in simulation & 0.1 -- 511 keV\\
Size of primary emission region, $r_{\rm er}$ & $8.8 \times 10^{13}$~cm ($6r_{\rm g}$)\\ 
Inner radius of simulated region, $r_{\rm min}$ & $4.4 \times 10^{16}$~cm \\
Outer radius of simulated region, $r_{\rm max}$ & $2.2 \times 10^{19}$~cm \\
Size of Cartesian Rad. Trans. simulation grid & $2.5 \times 10^{19}$~cm \\
3D Cartesian Rad. Trans. grid cells & $256\times256\times256$ \\
\hline 
\end{tabular}\\
\end{table*}

\begin{figure*}
\epsfig{file=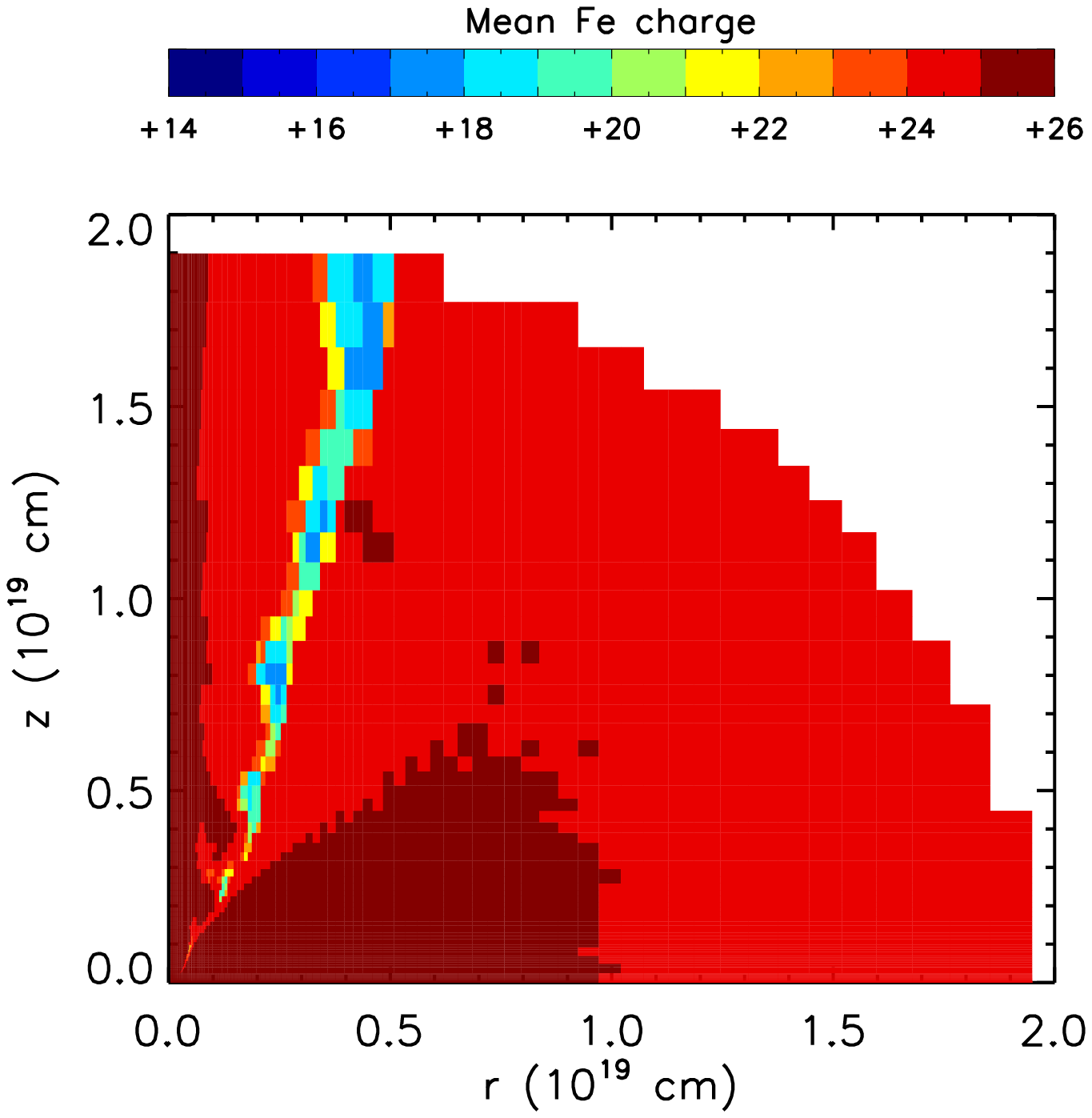,width=7cm,angle=90}
\epsfig{file=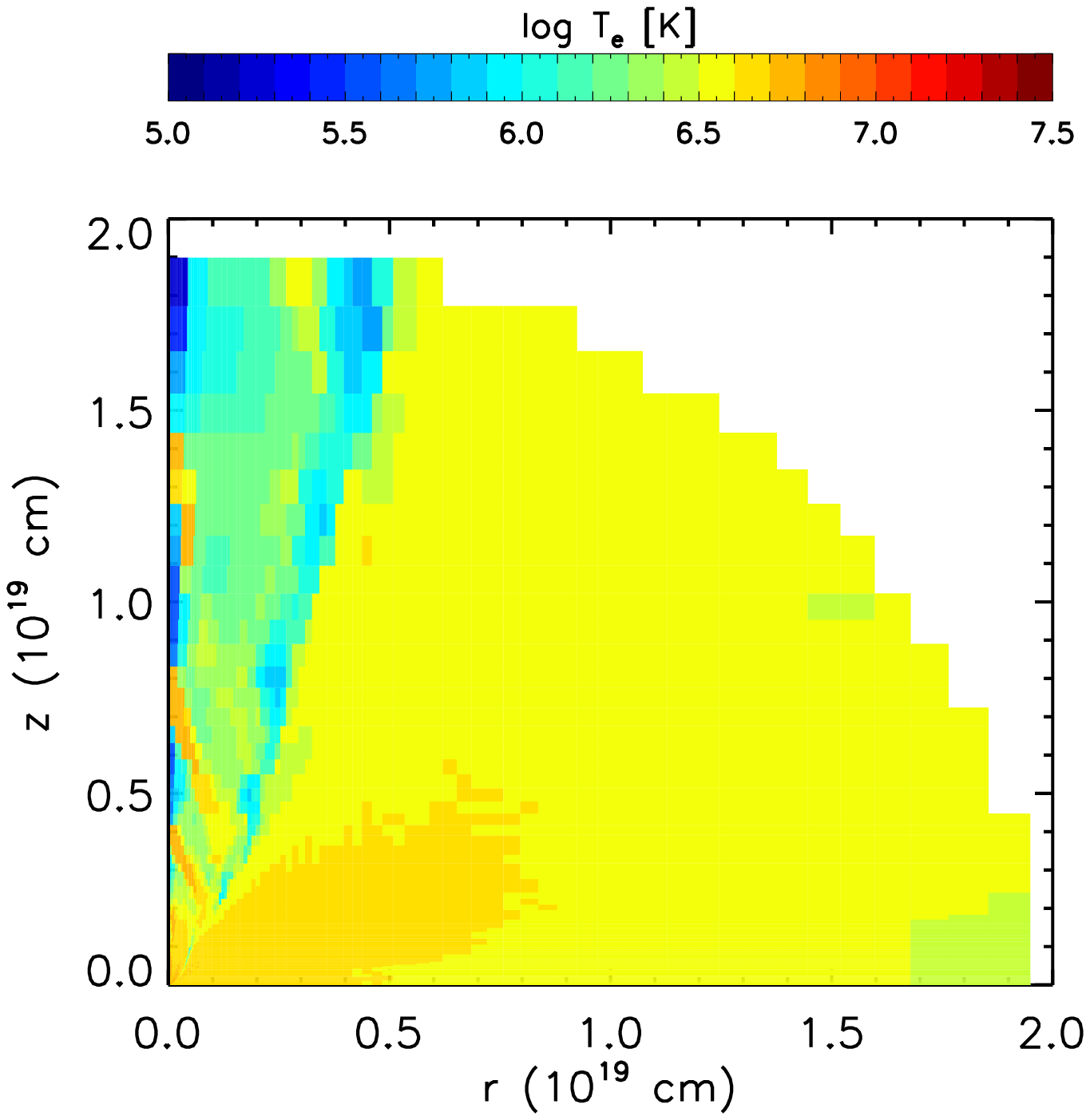,width=7cm,angle=90}
\epsfig{file=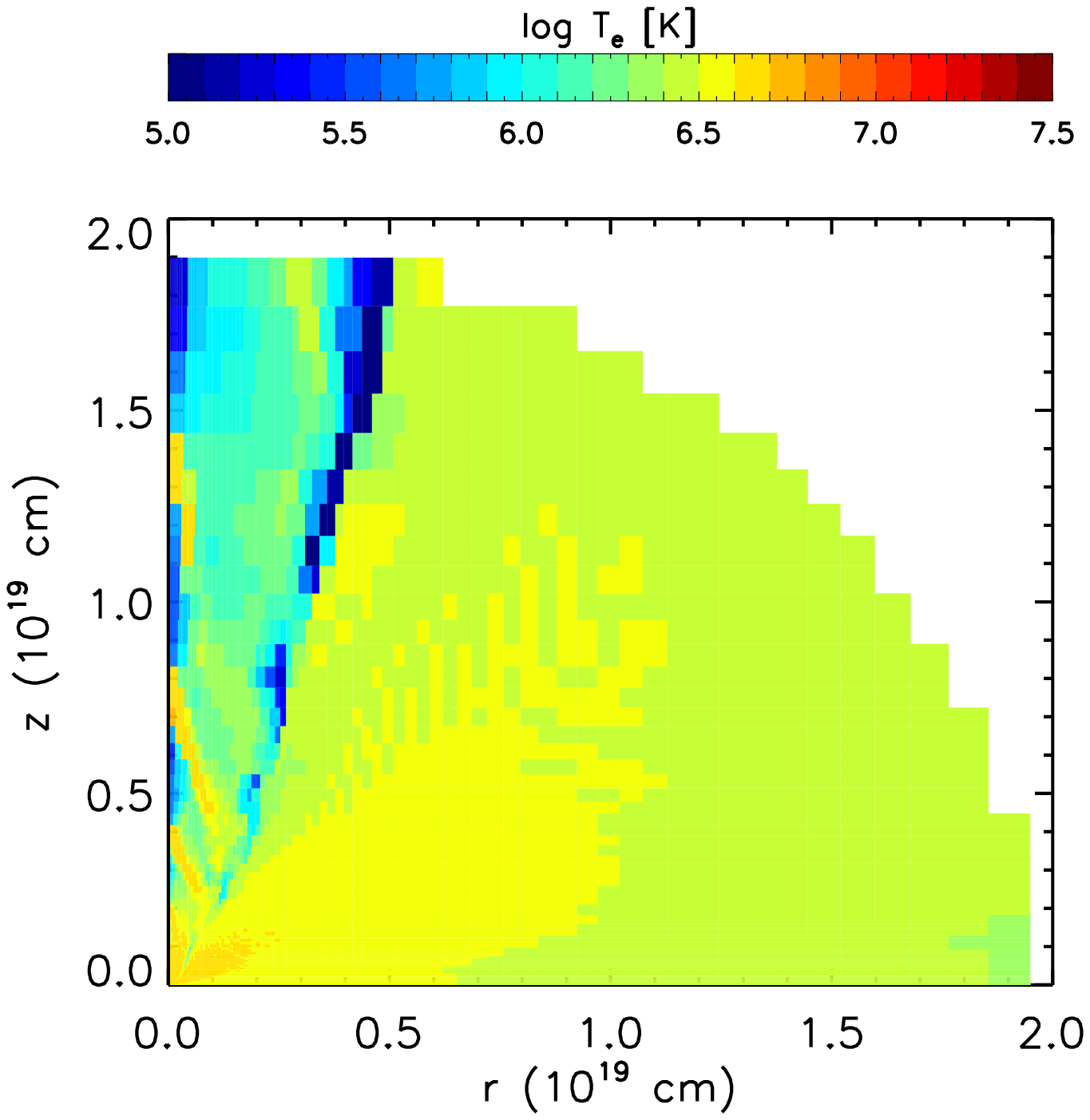,width=7cm,angle=90}
\epsfig{file=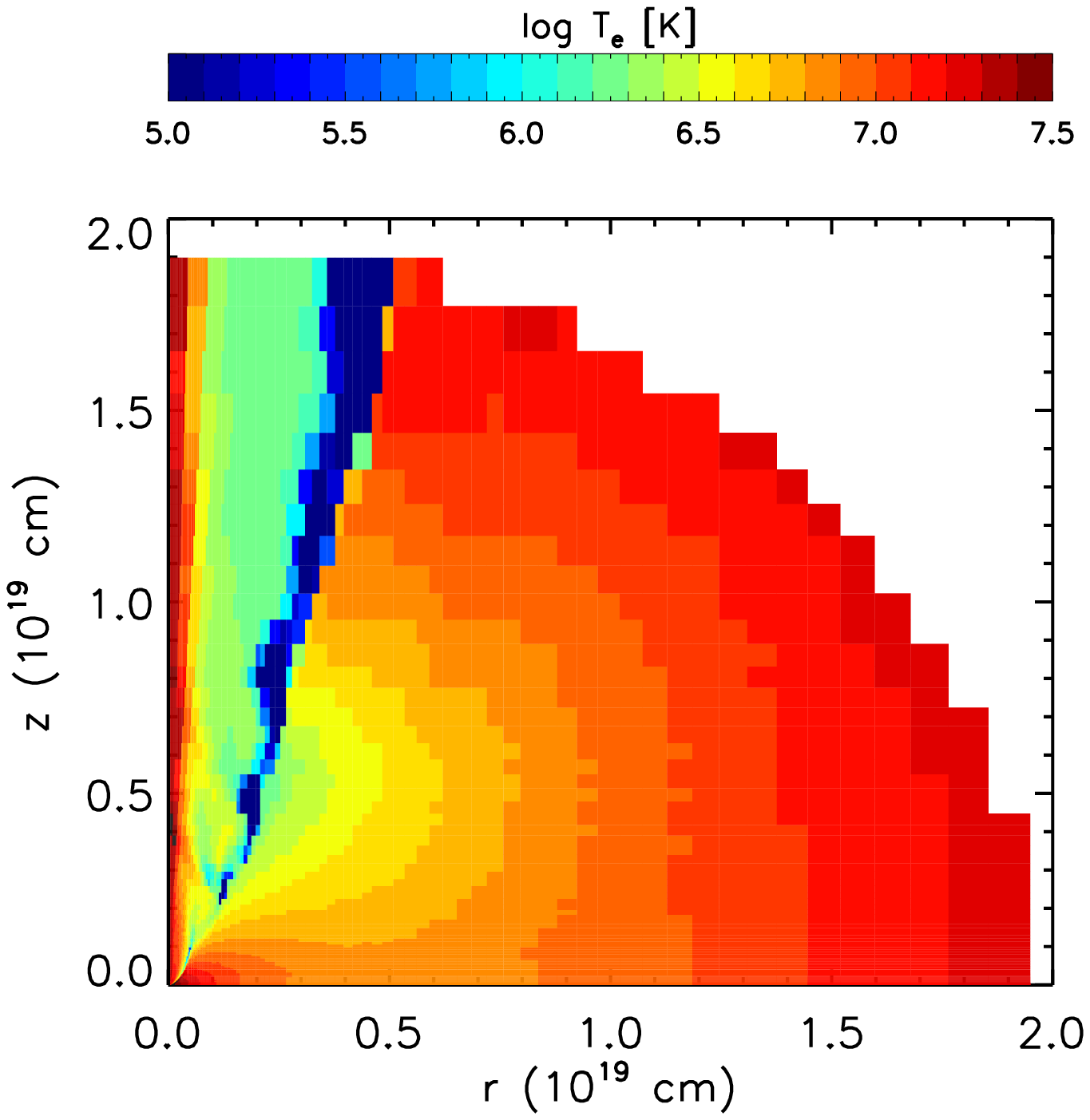,width=7cm,angle=90}
\caption{Mean charge of Fe ions (upper left) and the kinetic temperature (upper right) obtained from our radiative transfer simulations for the KP09-II model. The lower left panel shows the kinetic temperature we calculate if scattered light is neglected and the 
lower right panel shows the temperature originally computed by KP09.}
\label{fig:ionstate}
\end{figure*}

Table~\ref{tab:parameters} lists the important parameters adopted for the radiative transfer calculations (the same radiation parameters were adopted for our simulations of both models KP09-II and KP09-IV). Although our Monte Carlo method does not rely on a pre-defined photon frequency-grid, the effective resolution of the synthetic spectrum is limited by Monte Carlo noise to roughly $c \; {\Delta}E / E = 1000$~km~s$^{-1}$
in the simulations presented here.

As in \cite{sim10c}, we have assumed that the primary X-ray source is centrally located and isotropically emits a power-law spectrum with photon index $\Gamma = 2.1$. The source X-ray luminosity ($L_{X} = 1.5 \times 10^{44}$~ergs~s$^{-1}$) is chosen so that we match the flux assumed by KP09 in the 1 -- 20~keV region. 

Since the circularization radius for the infalling gas chosen by KP09 ($r_{\rm c} = 1800 r_{\rm g}$) is smaller
than the radius of the inner boundary of the simulation ($r_{\rm i} = 3000 r_{\rm g}$), an optically thick disk does not form in 
the computational domain.
Consequently, in contrast to \cite{sim10c}, we do not halt the propagation of 
our Monte Carlo quanta if they reach the $z=0$ plane but allow them to cross
unhindered.

\section{Results}
\label{sect:results}

\subsection{2D simulation (KP09-II)}
\label{sect:2D-results}

Figure~\ref{fig:ionstate} shows the mean ionization state of iron and the kinetic temperature we compute 
for the snapshot of the KP09-II model. 
Throughout most of the simulation volume, the gas is very highly ionized (the Fe ionization balance is dominated by Fe~{\sc xxv}, {\sc xxvi} and {\sc xxvii}) and the temperature is very close to the Compton temperature of our source radiation field ($T_{c} \sim 5 \times 10^6$~K). Only in the dense outflowing funnel of gas is the ionization (and temperature) reduced. Even in that region, however, the dominant Fe ion is no less ionized than Fe~{\sc xvii} and the temperature remains around $10^6$~K, the regime in 
which the gas is expected to be thermally unstable and fragment as found by KP09.
Overall, the degree of variation in ionization and temperature is significantly less than in the disk wind simulation considered by \cite{sim10c}, a consequence of the bright primary X-ray source and the absence of any very Compton-thick lines of sight through the KP09 simulations.  

Compared to KP09 (see Figure~\ref{fig:ionstate}), our temperatures are lower in the regions of infalling gas for two reasons.
First, adiabatic heating is included in the KP09 simulations but neglected here. Moreover,
our different assumptions about the spectrum of the primary radiation source mean that the Compton temperature in KP09 ($T_{c}  \sim 2 \times 10^{7}$K) is higher than ours ($T_{c}  \sim 5 \times 10^{6}$~K). Thus, our calculations largely agree that most of the 
volume of infalling gas is
heated to around the Compton temperature and the differences are mostly due to the assumed source spectral energy distribution.
This difference does not significantly impact our conclusions since even at our Compton temperature, the infalling gas is too
hot and ionized to imprint strong spectral features (see below).

In the dense clouds embedded in the outflowing gas we find higher temperatures than KP09. Additional 
radiative heating occurs in our simulations because of scattered light, which was not treated by KP09. To quantify the effect of scattered
light, we carried out an additional calculation in which scattered light was not included -- the temperature distribution obtained
is shown in Figure~\ref{fig:ionstate}.
Although less crucial here than in our disk wind simulations \citep{sim10c}, we find that scattered light is still important in
heating and ionizing gas that is partially shielded from direct irradiation by the central source. Further hydrodynamical simulations 
are needed to quantify how this influences the densities and velocities of the outflowing gas (see Section~\ref{sect:summary} for further discussion).

\begin{figure}
\epsfig{file=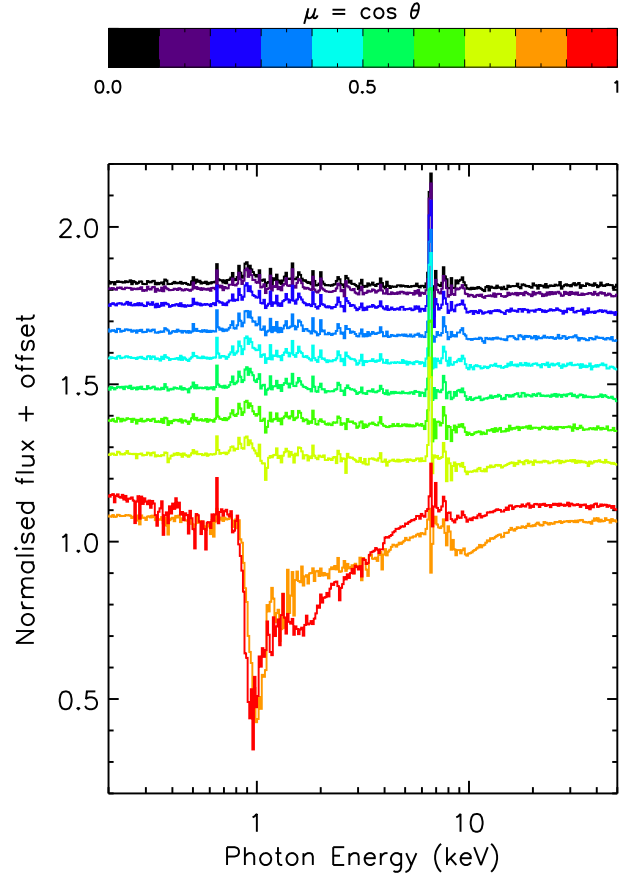,width=9cm,angle=0}
\caption{Computed spectra for the KP09-II model. Each line represents a different observer inclination angle ($\theta = \cos^{-1} \mu$) to the polar axis. All spectra are normalized to the primary X-ray power-law spectrum and are
shown at a spectral resolution $c \; {\Delta}E / E = 5000$~km~s$^{-1}$.
For clarity, the spectra are offset from each other by additive constants (the $\mu = 1$ spectrum is shown with no offset; each of the others is offset by +0.1, relative to the spectrum for the next higher $\mu$-value).}
\label{fig:spec_2d_all}
\end{figure}

We have computed spectra for the KP09-II model for a set of twenty observer orientations. These uniformly sample the range of inclinations $0 < \mu < 1$, where $\theta = \cos^{-1} \mu$ is the angle between the observer's line-of-sight and the polar axis of the model. A sub-set of these spectra
are shown in Figure~\ref{fig:spec_2d_all}.

\begin{figure*}
\epsfig{file=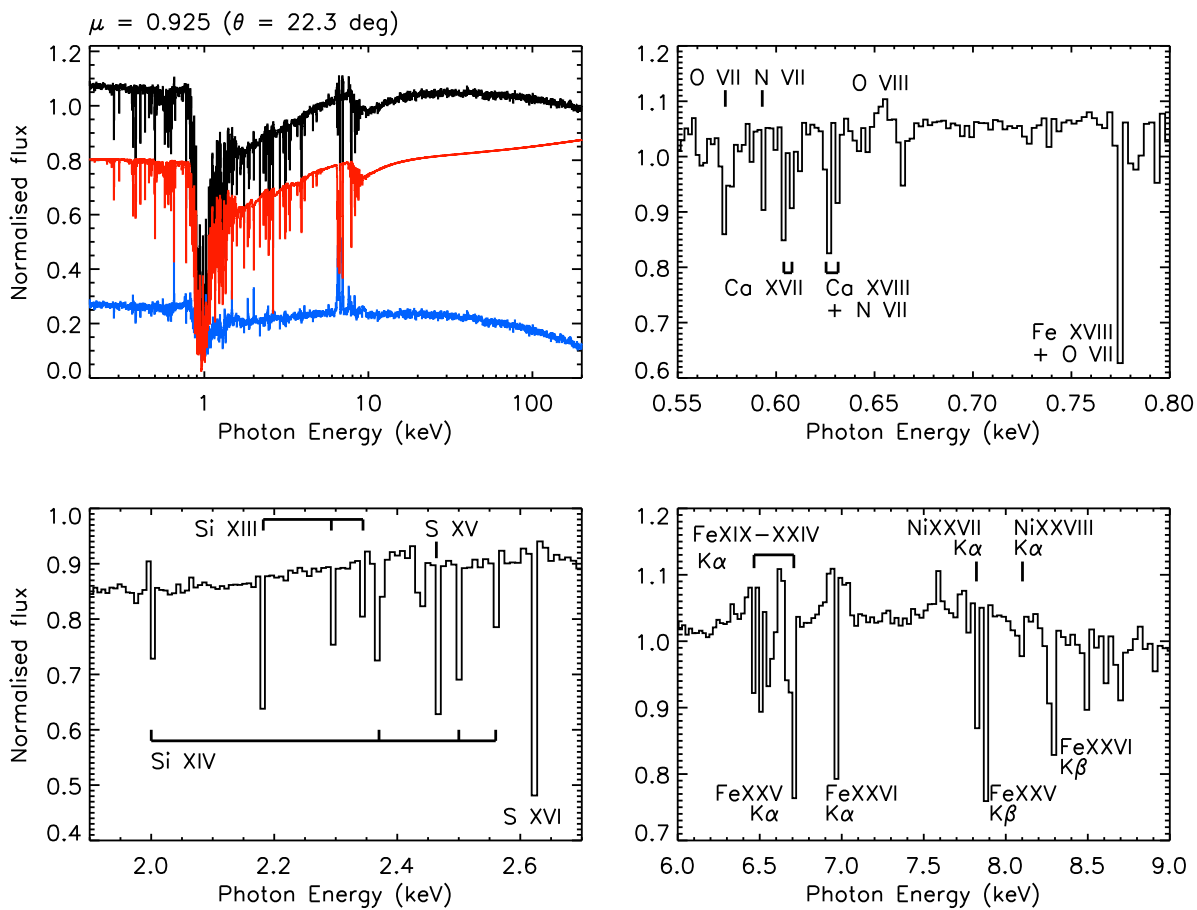,width=16cm,angle=0}
\caption{Computed spectra for the KP09-II model for an observer inclination angle of $\mu = \cos\theta = 0.925$. In the top left panel, the black line shows the synthetic spectrum. The red and blue lines show the contributions from photons reaching the observer directly from the primary X-ray source and following scattering/reprocessing in the flow, respectively. The remaining panels show details of the spectral regions around the O~{\sc viii}, Si~{\sc xiv} and  Fe~K$\alpha$ features. The spectrum is normalized to the primary X-ray power-law spectrum and shown at a spectral resolution corresponding to $c \; {\Delta}E / E = 1000$~km~s$^{-1}$. Monte Carlo noise is responsible for percentage-level fluctuations in the spectra.}
\label{fig:spec_2d}
\end{figure*}

The synthetic spectra computed for the KP09-II model roughly divide into two classes, distinguished by the the observer orientation relative to the outflowing funnel of gas. As expected, the most complex spectra are predicted for observer orientations closer to the pole than the outer boundary of the funnel ($\theta \simlt 35$~deg). Figure~\ref{fig:spec_2d} show an example of this type of spectrum in detail. Roughly speaking, the spectrum consists of two components: the direct spectrum of photons from the the primary X-ray source and a contribution of scattered light generated 
throughout the simulation volume (these components are shown separately in the top left panel of Figure~\ref{fig:spec_2d}). 
The relatively dense gas in the polar funnel imprints strong photoelectric and line absorption features in the direct spectrum of the primary source. In contrast, the spectrum of scattered light shows only relatively weak absorption since it is predominantly formed in the very high ionization gas located outside the polar funnel. However, scattering and fluorescence in the outflow do give rise to a range of weak emission features, particularly in the Fe~K region.

\begin{figure*}
\epsfig{file=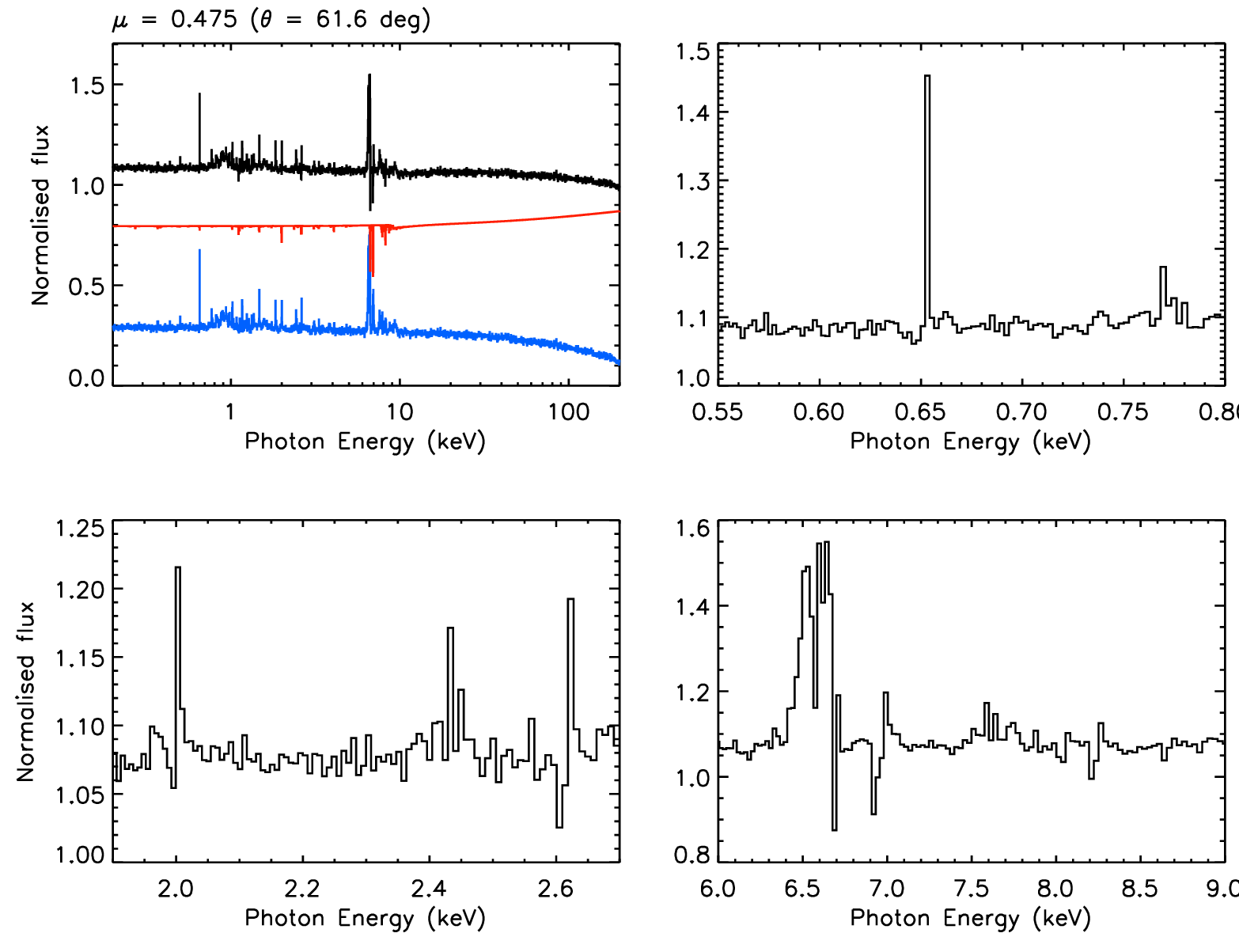,width=16cm,angle=0}
\caption{As Figure~6, but for an observer inclination angle of $\mu = \cos\theta = 0.475$.}
\label{fig:spec_2d2}
\end{figure*}

For higher inclination angles ($\theta \simgt 35$~deg), the gas in the observer line-of-sight is too highly ionized for photoelectric absorption to affect the spectrum significantly (see Figure~\ref{fig:spec_2d2}). 
For these orientation, the strongest spectral features are the emission lines produced by fluorescence in the polar funnel region. In particular, there is noticeable net emission in the Fe~K region ($\sim 6.5$~keV, a blend of individual lines from the K-shell and high L-shell ions). There are a few very weak absorption features in the spectrum, associated with the very highly-ionized gas that dominates the volume outside the polar funnel. In particular, the Fe~{\sc xxvi} K$\alpha$ line appears as a clear {\it inverse} (but weak) P Cygni profile (see lower right panel of Figure~\ref{fig:spec_2d}), arising from the net inflow in most of the computational grid. Even weaker, but similar, features are also predicted in e.g. S~{\sc xvi} K$\alpha$ and Si~{\sc xiv} K$\alpha$ (lower left panel of Figure~\ref{fig:spec_2d2}).

Since the flow velocities in the KP09 simulations are fairly small (at most a few thousand km~s$^{-1}$ but more typically a few hundred km~s$^{-1}$), all the Fe~K line features in the model spectra are relatively narrow such that details of the profiles are not well-resolved by current instrumentation (e.g. the {\it Chandra} High Energy Transmission Grating has a resolution full-width-at-half-maximum of $\sim 1800$~km~s$^{-1}$ around the Fe~K region). 
However,
the photoelectric absorption due to the polar funnel material (predicted at low inclination) and the net Fe~K$\alpha$ emission (strongest at moderate to high inclination) should be detectable.
The absence of strong absorption signatures associated with the infalling gas is an 
important result. It implies that that the observed lack of red-shifted absorption
does not exclude the inflow-outflow pattern of the KP09 (and similar) simulations.

\subsection{3D model (KP09-IV)}

The polar symmetry imposed in the 2D model inhibits the development of substructure in the polar funnel. As noted in Section~\ref{sect:sims}, in the 3D simulation (KP09-IV) the material in the funnel region fragments into clumps and clouds, leading to significant variations in the column density through the flow for different azimuthal orientations (see Figure~\ref{fig:columns_phi}). Here our goal is to quantify how this additional variation affects the spectral features.

To this end, we computed a total of 100 synthetic spectra for model KP09-IV. These explore five values of the azimuthal angle ($\phi = 30$, 60, 90, 120 and 150 degrees) and the same set of twenty $\mu$-values used for the KP09-II spectra.
These KP09-IV spectra are generally similar to those calculated from KP09-II and can again be subdivided into two broad categories: spectra with strong photoelectric absorption (for observer orientations relatively close to the pole) and those in which the only significant spectral features are narrow fluorescence emission lines.

\begin{figure*}
\epsfig{file=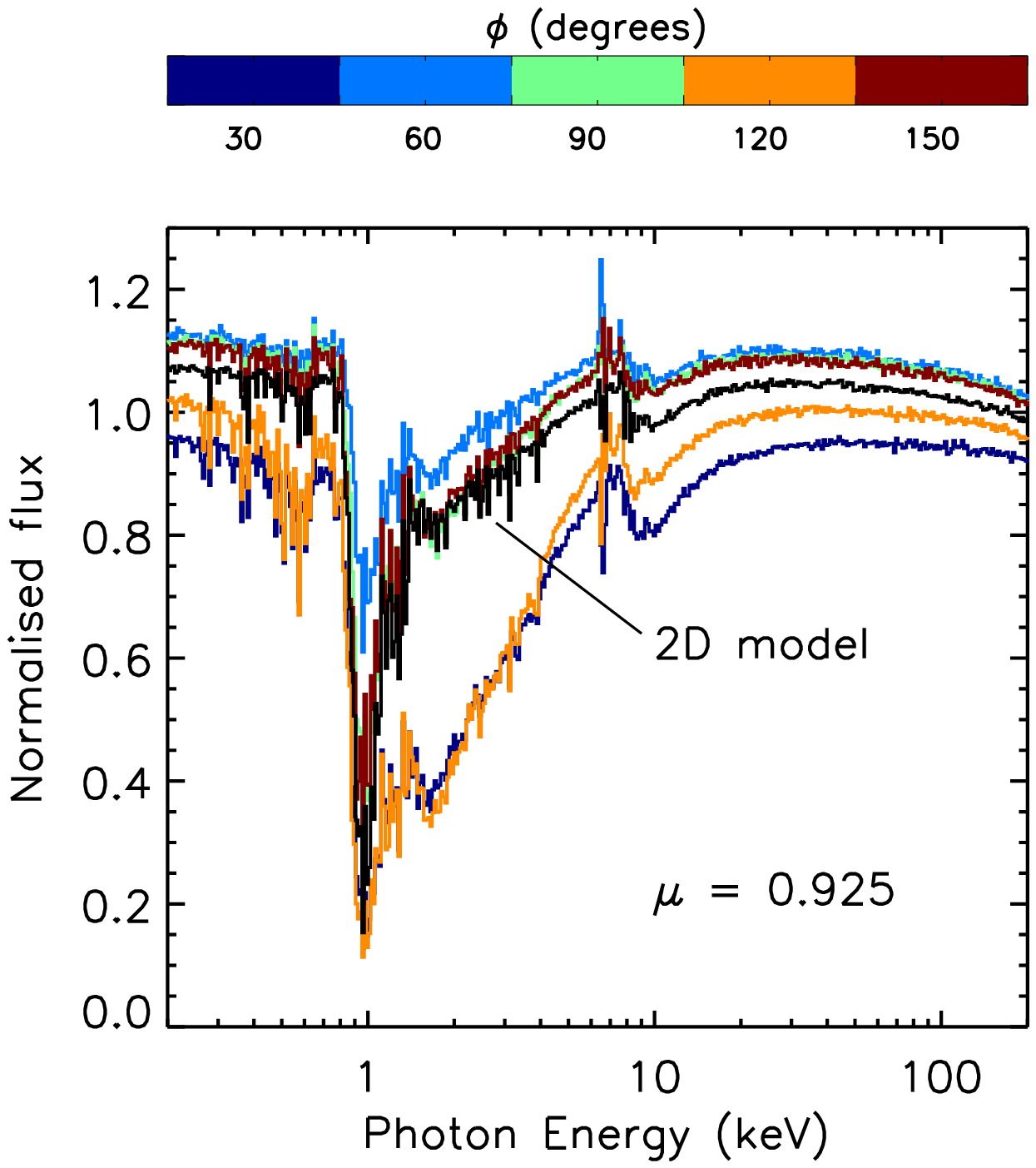,width=8cm,angle=0}
\epsfig{file=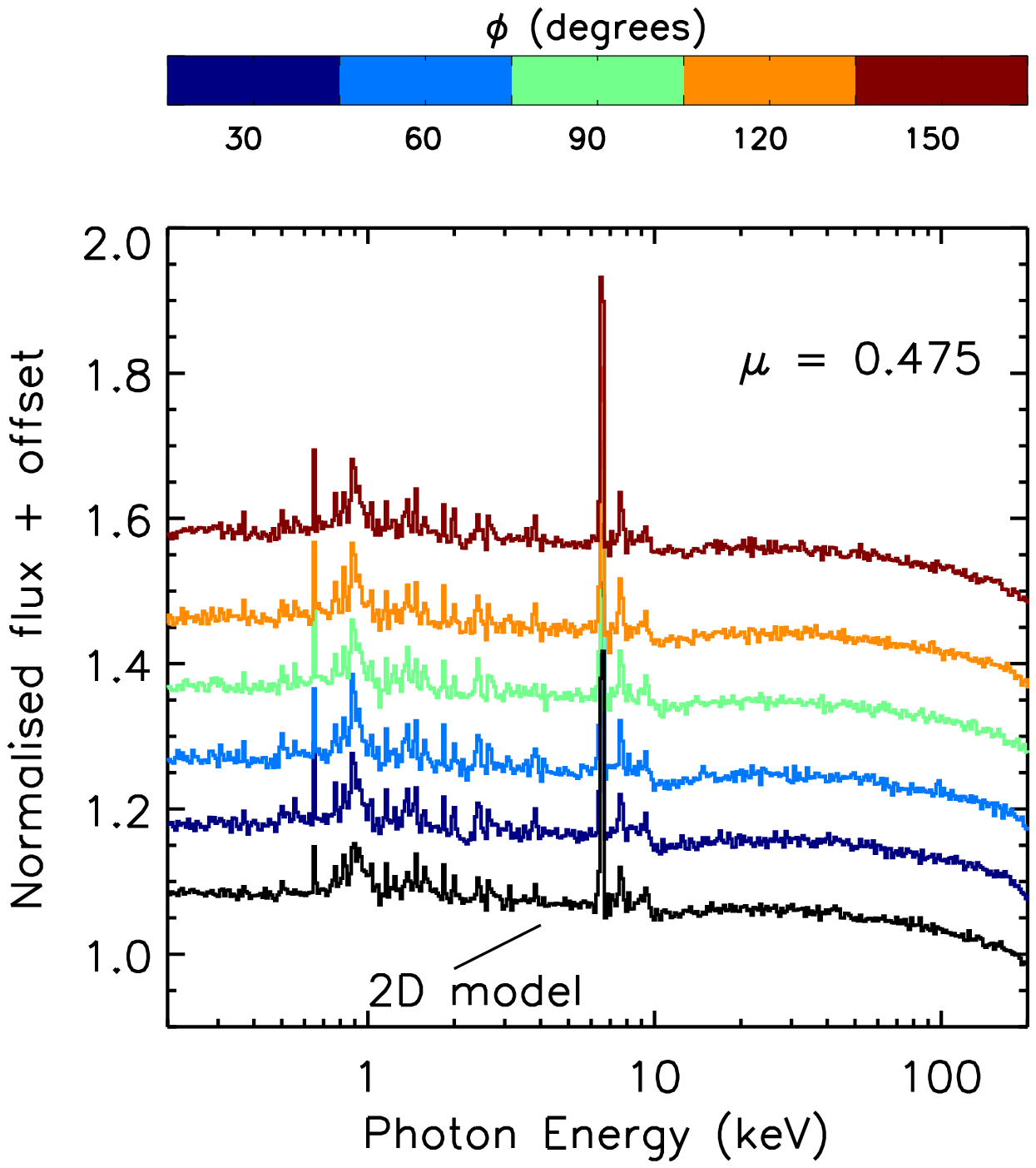,width=8cm,angle=0}
\caption{Comparison of computed spectra for the KP09-IV (3D) and KP09-II (2D) models for $\mu = 0.925$ (left) and $\mu = 0.475$ (right). In both panels, the KP09-II spectra are shown in black while KP09-IV spectra for five differing $\phi$-values are shown by the coloured lines. 
All spectra are normalized to the primary X-ray power-law spectrum and are
shown at a spectral resolution $c \; {\Delta}E / E = 5000$~km~s$^{-1}$.
In the right panel, the KP09-IV spectra have been offset by additive constants.}
\label{fig:spec_3d}
\end{figure*}

As expected (see Figure~\ref{fig:columns_phi}), the 3D flow structure significantly increases the diversity of absorption features that can appear
in the spectrum for near polar inclinations. This is illustrated in Figure~\ref{fig:spec_3d}, which compares the synthetic spectra from the KP09-IV simulation to the KP09-II model at $\mu = 0.925$. However, for observer orientations away from the polar axis (i.e. cases in which there is no strong photoelectric absorption), the differences between the KP09-II and KP09-IV spectra are minor (right panel of Figure~\ref{fig:spec_3d}), indicating that the extra structure present in the 3D simulation does not strongly influence the fluorescent emission from the polar region. This is expected since the emission is formed as an integral over
the complete flow and neither the total mass outflow rate not the mean volume occupied by the
flow are very different between the KP09-II and KP09-IV simulations.

\section{Discussion}
\label{sect:discuss}

Our calculations shows that the KP09 models predict both absorption and emission features that could be detectable in X-ray spectra. 
Below, we comment on the comparison 
of these features with observations of AGN. 

\subsection{Absorption features}

Most of the spectral features are associated with the outflowing polar funnel of gas in the KP09 simulations. For observer orientations where the funnel is close to the line-of-sight ($\theta \simlt 35$~deg.), it causes significant photoelectric absorption in the 1 -- 10~keV band and also imprints a forest of narrow, low-velocity absorption lines associated with moderate-to-high ionization gas.

The absorption components predicted in our simulations are not expected to be strongly variable on timescales typically probed by X-ray observatories 
 -- the simulations cover a region extending from $\sim 0.01$ to $7$~pc and have gas velocities of at most a few thousand km~s$^{-1}$. Therefore the timescale for significant variations is expected to be greater than the decade-long timescale of modern X-ray missions. Some degree of variation in the degree of ionization may be possible (driven by changes in the primary X-ray source properties) but this is expected to be small and washed out by the long recombination times associated with low-density material at large distances \citep[see e.g.][]{kaastra12}. 

Therefore, the absorption predicted in our simulations is unlikely to be physically related to observed absorption components that manifest rapid variability. However, it may be related to persistently observed absorption systems that do not show strong variability and are consequently thought to be relatively distant from the AGN. Such absorption components have been identified in a number of sources.
For example, \cite{kaastra12} used the lack of short-term variability to 
constrain the location of several ionized absorption components in the X-ray spectra of Mrk 509 to be $> 5$pc from the central black hole.
Similarly, \cite{krongold10} constrained the gas responsible for 
a highly-ionized absorption component in NGC~5548 to lie at $> 0.03$pc while
\cite{behar03} and \cite{netzer03} favour distances of $> 0.18 - 10$pc for the
main warm absorbers in NGC~3783 (although we note that \citealt{reeves04} argue 
that the most highly-ionized absorption in NGC~3783 may be located at smaller 
radii, $\simlt 0.1$~pc). 
These 
distance limits are roughly consistent with the location of the gas in the KP09 simulations. Also, the observationally inferred outflow velocities are comparable to those in the theoretical models ($\simlt 1000$~km~s$^{-1}$). 

However, the total column density associated with the ionized absorption by the polar funnels in our simulations and the degree of ionization are typically higher than that inferred from the observations. For Mrk 509, Kaastra et al. (2012) measure hydrogen column densities for their most highly ionized absorption components (D and E) of only $\sim 6 \times 10^{20}$~cm$^{-2}$ and $60 \times 10^{20}$~cm$^{-2}$ while values up to $\sim 10^{22}$~cm$^{-2}$ have been suggested for both NGC~3783 and NGC~5548 \citep{behar03,netzer03,krongold10}.
In contrast,
the KP09 column densities are typically $\sim 10^{23}$~cm$^{-2}$ in the polar region, making the corresponding component of ionized absorption too strong. In addition, observations show clear evidence of significant absorption by lower ionization gas than is present in the simulations (e.g. the unresolved transition arrays of M-shell Fe ions; see \citealt{behar01}). These discrepancies clearly point to quantitative shortcomings of
the properties of the outflow gas in the simulations and motivate further exploration of the model parameter space 
(see Section~\ref{sect:summary} for further discussion).

Our calculations predict no strong absorption 
features associated with the infalling gas, despite the significant column densities 
associated with this component of the flow (see Figure~\ref{fig:columns}). Thus, the observed
absence of strong red-shifted absorption features 
in X-ray spectra does not exclude
models for outflows driven from inflows. Apart from the weak Fe~{\sc xxvi} inverse P Cygni line (Section~\ref{sect:2D-results}), the main effect of the infalling gas is to
generating a smooth continuum of scattered light (see Section~\ref{sect:discuss-sc}).

\subsection{Emission features}

Scattering and fluorescence in the outflowing gas gives rise to emission lines, including significant Fe~K$\alpha$. These emission features appear for all observer orientations and are the only strong feature for inclination angles larger than the opening angle of the polar funnel. The equivalent width of the Fe~K$\alpha$ emission is $\sim 100$~eV for most observer orientations. This is comparable to the strength of the narrow K$\alpha$ line cores observed in Seyfert galaxies \citep{yaqoob04,shu10}, suggesting that the amount of fluorescing gas in the KP09 models may be roughly appropriate. However, the position and structure of the synthetic line features are not consistent with typically observed narrow lines. In particular, \cite{yaqoob04} and \cite{shu10} find lines with energy close to 6.4~keV and full-width-at-half-maximum (FWHM)
of $\sim 2000$~km~s$^{-1}$. Although the
velocities in our models are roughly consistent with this line width ($\simlt 1000$~km~s$^{-1}$), our synthetic spectra contain multiple line components due to 
the range of ionization states present in the outflowing gas (see Figures~\ref{fig:spec_2d} and \ref{fig:spec_2d2}). These different
line components could be marginally resolved in high-quality data and cause the
Fe~K$\alpha$ emission complex to extend over a much wider 
spectral region (equivalent to 
$\sim 10,000$~km~s$^{-1}$) than is typically observed. 
Furthermore, the flux is concentrated around 
6.5 -- 6.6~keV rather than 6.4~keV. Thus, compared
to the commonly observed properties of narrow Fe~K emission, it is clear 
that our calculations produce too much emission from significantly ionized 
material (Fe~{\sc xix} -- {\sc xxiv}).

\begin{figure}
\epsfig{file=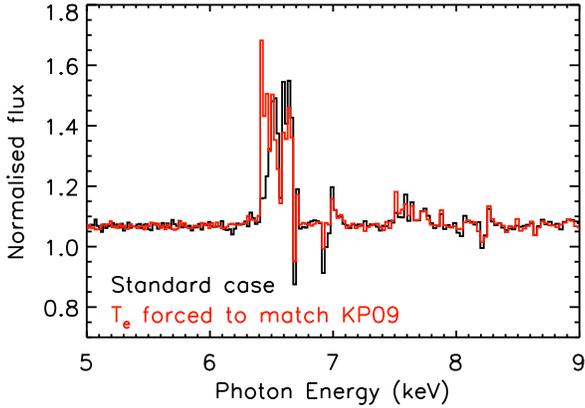,width=8cm,angle=0}
\caption{Computed Fe~K spectra at $\mu = 0.475$ for the KP09-II model adopting the kinetic temperatures from KP09. For comparison, 
we also show the results from our standard calculation (see Figure 7).
All spectra are normalized to the primary X-ray power-law spectrum and shown at a spectral resolution corresponding to $c \; {\Delta}E / E = 1000$~km~s$^{-1}$.}
\label{fig:spec2d_fixT}
\end{figure}

Given that our calculations predict that the Fe~K$\alpha$ emission should appear for the majority of observer orientations, the discrepancies in spectral structure in the region around 6.4 -- 6.7~keV suggest that the conditions predicted in our current simulations are not typical of Type~I AGN. However, we note that the discrepancies could be largely removed if the typical ionization state of the material in the outflowing gas were reduced. To qualitatively illustrate the sensitivity of the Fe~K$\alpha$ emission to the flow conditions, we have also
computed synthetic spectra for the 2D KP09-II model in which we fixed the temperature to that obtained by KP09 (as discussed in Section~\ref{sect:2D-results}, they generally found lower temperatures in the dense regions of the outflow, which result in reduced ionization compared to our full radiative transfer calculation). The synthetic spectrum from this calculation is compared to our standard case in Figure~\ref{fig:spec2d_fixT}. 
This shows that the K$\alpha$ emission is affected by details of the conditions in the outflow -- adopting the KP09 temperatures changes the ionization state sufficiently that more of the flux emerges around 6.4~keV.
However, this change alone does not qualitatively alter our findings -- a more
significant reduction of the ionization state would be required to shift
most of the K$\alpha$ emission to 6.4~keV. This might be achievable via
stronger clumping in the outflow, as will be discussed in Section~\ref{sect:summary}.

We note that our models also predict an emission component from Fe~{\sc xxvi} around 6.9~keV, which is well-separated from the main Fe~K$\alpha$ emission in our spectra (6.4 -- 6.7~keV). Narrow Fe~{\sc xxvi} emission components
have been reported in several objects (e.g. NGC~7314 [\citealt{yaqoob03}]; NGC~7213 [\citealt{bianchi08}; NGC~3516 [\citealt{turner08}]) and based on a sample 
of {\it Suzaku} spectra, \cite{patrick11} suggested that such a feature is 
relatively common 
(although we note that this result is somewhat dependent on the modelling, 
see \citealt{tatum12}).

Particularly for inclination angles beyond the polar funnel ($\theta \simgt 35$~deg.), our synthetic spectra also contain emission lines
associated with H-like ions of lighter elements, most prominently O~{\sc viii} Lyman~$\alpha$ (but also S~{\sc xvi}, Si~{\sc xiv} and Mg~{\sc xii} Lyman~$\alpha$). Such H-like emission lines have been reported in e.g. the spectra of the Seyfert 2 galaxy NGC 1068 \citep{kinkhabwala02,ogle03}. This is qualitatively consistent with our simulations, which predict that emission lines are the only strong spectral features imprinted for relatively high inclinations.
However, there are important differences. In particular, the observations also include many lines of lower ionization material (e.g. \citealt{kinkhabwala02} and \citealt{ogle03} report strong emission in both O~{\sc vii} and O~{\sc viii} while the model spectra show virtually no O~{\sc vii} emission lines). There is also a kinematic mismatch -- e.g. the synthetic O~{\sc viii} line profile is not resolved in our Monte Carlo spectra (FWHM $< 500$~km~s$^{-1}$) while the observed profile is broader (FWHM $\sim 1500$~km~s$^{-1}$), suggesting that higher fluid velocities are required. 

To summarise, comparison of the emission line properties leads to similar conclusions to those drawn from the absorption features: the characteristic properties of the calculated spectra are encouraging but there are clear quantitative shortcomings. In particular, the ionization state is generally too high in the simulations.

\subsection{Scattered continuum}
\label{sect:discuss-sc}

In all our synthetic spectra, a significant fraction of the predicted continuum flux ($> 10$ per cent; see the scattered light component in Figures~\ref{fig:spec_2d} and \ref{fig:spec_2d2}) is produced by Compton scattering in the highly ionized gas outside the polar funnel. 
Consequently, even for orientations for which the absorbing column through the funnel is very high, the observer will see a component of weakly attenuated continuum flux (produced by Compton scattering into the line-of-sight). As a result, absorption features are not expected to reach zero flux, despite high optical depths. The spectrum can therefore resemble that of a partial-covering model (in which the spectrum is described by a linear combination of components, typically one unabsorbed continuum component plus a significantly absorbed component). 
An example of scattered continuum re-filling absorption lines, albeit on a different scale (~150 pc), has been seen in the case of 
Mrk~509 \citep{kriss11}.

\section{Summary and future work}
\label{sect:summary}

We have calculated synthetic X-ray spectra for the KP09 two- and three-dimensional simulations of outflows driven from infalling gas close to an AGN. We showed that the polar funnel of outflowing gas predicted by the KP09 simulations has observable consequences for all observer orientations. It causes strong, ionized absorption features for near face-on inclinations and ionized fluorescent emission for all lines of sight.
The shapes of the spectral features predicted for our two- and three-dimensional simulations are generally similar. However, in the three-dimensional case, we do find significant variation (a factor of $\sim 3$) in the depth of absorption with azimuthal angle (see Figure~\ref{fig:spec_3d}), which is a consequence of the clouds/clumps in the flow.

Conversely, we found that the infalling gas, which occupies most of the volume outside the polar funnel in the simulations, does not imprint strong spectroscopic absorption features because of the very high ionization state.
Only a weak Fe~{\sc xxvi} K$\alpha$ feature is predicted
and we note that our simulations may even over-estimate the strength of this feature (non-radiative heating likely 
makes the temperature of the infalling gas even higher; see Section~\ref{sect:2D-results}).
Thus we conclude that an observed lack of inflow features does not rule out outflow-from-inflow models similar to those presented by KP09. 

When compared to the properties of observed absorption and emission features associated with gas that is relatively distant from the black hole, the particular simulations presented here face several important challenges. Although the outflow velocities are roughly consistent with observations,
the column density and typically ionization state of the outflowing gas are too high in the models.

The ionization state in our simulations could be lowered by simply reducing the source X-ray luminosity. However, the reduction required (a factor of $\sim 10$) would be quite inconsistent with the properties of the radiation source assumed in the KP09 hydrodynamical simulations. Moreover, reduced luminosity leads to a global reduction in the ionization of all the gas -- consequently, the infalling gas (occupying $\theta \simgt 35$~deg) would cease to be fully ionized and start to imprint spectral features.
The fact that the absorption predicted for polar orientations ($\theta \simlt 35$~deg) is also too strong in the simulations makes this difficulty harder to overcome. If the flow pattern remains the same, the gas density would need to be lower to give a smaller column density but a reduced density would make the over-ionization problem more severe.

More promising than any global change of the density or X-ray flux might be smaller scale changes in the structure of the outflowing gas in the polar 
funnel. For a similar total mass outflow rate,
higher typical densities in the outflow might be achieved 
if it were more strongly clumped. This would not only lead to lower ionization but could also reduce the column density since smaller clumps will subtend smaller solid angles as seen by the central source.

It is important to note that the KP09 simulations, and the associated radiative transfer calculations presented 
here, are only initial, exploratory results. A major conclusion to be drawn
is that thermal instabilities are likely to cause the flow to fragment, which has observable consequences -- that the thermal instability may be at work in AGN outflows is already supported by spectroscopic evidence \citep{holczer07}.
More detailed work is now needed to explore the parameter space of the simulations and investigate 
physical and numerical aspects of the fragmentation process.
For example, here we have considered only one case for the angular momentum of the infalling gas 
(corresponding to a circularization radius of $r_{\rm c} = 1800 r_{\rm g}$). Changing the angular
momentum will alter the relative densities of polar and equatorial parts of the flow.
Moreover, although the KP09 simulations do show that the outflowing gas breaks into clouds, the scale of fragmentation is limited by the resolution of the simulations and depends on factors such as the accuracy of the opacities used in the hydrodynamical simulations. 
Further investigation of fragmentation in simulations for gas infalling towards an AGN have been made by
\cite{barai11,barai12}. 
They carried out simulations extending to 
larger spatial scales than KP09 but did not include the radiative line-force.
Their results confirm that 
thermal instabilities can cause clumps to form and, depending 
on simulation parameters, that this
can also happen in the infalling gas.
How such fragmentation of the inflow might affect the dynamics of a radiatively driven outflow and the
observables now needs to be
investigated via simulations similar to KP09 with extended computational domains.
In addition, our radiative transfer simulations have 
demonstrated that scattered light can be significant in heating and ionizing the outflowing material (see also \citealt{sim10c}), an effect
which is not currently incorporated in the hydrodynamical simulations. Simply adopting the temperature distribution derived by KP09
in our radiative transfer simulations does not qualitatively change our results (although our synthetic spectra are quantitatively affected). Nevertheless, hydrodynamical simulations that consider scattered light are needed since the additional heating may affect 
the onset of thermal instabilities and ultimately impact the density and velocity distribution of the outflowing gas.

We conclude that outflows driven from inflows are promising as a possible site for observable X-ray absorption and 
fluorescent emission around AGN. However, further study is required: in particular, the next step should be detailed, 
high-resolution studies of the clumping and structure of outflows and how these
depend on key simulation parameters.

\section*{Acknowledgments}

We acknowledge support provided by the Chandra award
TM0-11010X issued by the Chandra X-Ray Observatory Center,
which is operated by the Smithsonian Astrophysical Observatory for and on
behalf of NASA under contract NAS 8-39073.
DP also acknowledges the UNLV sabbatical assistance.
T.J.Turner acknowledges NASA grant NNX11AJ57G.
We thank the referee for constructive comments on our work.

\bibliographystyle{mn2e}
\bibliography{snoc}

\label{lastpage}
\end{document}